\documentclass[fleqn,usenatbib]{mnras}

\usepackage{newtxtext,newtxmath}

\usepackage[T1]{fontenc}
\usepackage{ae,aecompl}
\usepackage{ulem}

\usepackage{graphicx}	
\usepackage{amsmath}	

\newcommand\msunh{\,h^{-1}\rm{\,M_{\odot}}}
\newcommand\msun{\,\rm{\,M_{\odot}}}
\def\kms{\rm{\,km\,s^{-1}}}

\title[BCGs and centers]{Brightest Cluster Galaxies: the centre can(not?) hold}

\author[R. De Propris et al.]
{Roberto De Propris$^{1}$, \thanks{E-mail: rodepr@utu.fi}
Michael J. West$^{2}$,
Felipe Andrade-Santos$^{3}$,
\newauthor
Cinthia Ragone-Figueroa$^{4,5}$,
Elena Rasia$^{5,6}$,
William Forman$^{3}$,
Christine Jones$^{3}$,
\newauthor
Rain Kipper$^{7}$,
Stefano Borgani$^{5,6,8,9}$
Diego Garc\'ia Lambas$^{4}$,
Elena A. Romashkova$^{10}$ 
\newauthor
and Kishore C. Patra$^{11}$ \\
\\
$^{1}$FINCA, University of Turku, Vesilinnantie 5, Turku, 21400, Finland\\ 
$^{2}$Lowell Observatory, 1400 W Mars Hill Rd, Flagstaff, AZ 86001, USA \\
$^{3}$Harvard-Smithsonian Center for Astrophysics, 60 Garden Street, Cambridge, MA 02138, USA\\
$^{4}$ IATE, CONICET, Universidad Nacional de C\'ordoba, Laprida 854, X5000BGR, C\'ordoba, Argentina\\
$^{5}$ INAF, Osservatorio Astronomico di Trieste, via Tiepolo 11, I-34131, Trieste, Italy \\
$^{6}$ IFPU - Institute for Fundamental Physics of the Universe, Via Beirut 2, 34014 Trieste, Italy\\
$^{7}$ Tartu Observatory, University of Tartu, Observatooriumi 1, 61602 T\~oravere, Estonia\\
$^{8}$ Dipartimento di Fisica dell’ Universit`a di Trieste, Sezione di Astronomia, via Tiepolo 11, I-34131 Trieste, Italy\\
$^{9}$INFN - National Institute for Nuclear Physics, Via Valerio 2, I-34127 Trieste, Italy\\
$^{10}$Department of Physics, Massachusetts Institute of Technology, Cambridge, MA 02139, USA\\ 
$^{11}$Department of Astronomy, University of California, Berkeley, CA 94720-3411, USA
}

\date{Accepted XXX. Received YYY; in original form ZZZ}

\pubyear{2020}

\begin{document}
\label{firstpage}
\pagerange{\pageref{firstpage}--\pageref{lastpage}}
\maketitle

\begin{abstract}

We explore the persistence of the alignment
of brightest cluster galaxies (BCGs) with their
local environment. We find that a significant
fraction of BCGs do not coincide with the centroid of the X-ray gas distribution
and/or show peculiar velocities (they are not at rest with respect to the cluster mean). Despite
this, we find that BCGs are generally aligned
with the cluster mass distribution even when
they have significant offsets from the X-ray
centre and significant peculiar velocities.
The large offsets are not consistent with
simple theoretical models. To account for
these observations BCGs must undergo mergers
preferentially along their major axis, the 
main infall direction. Such BCGs may be
oscillating within the cluster potential after
having been displaced by mergers or collisions, 
or the dark matter halo itself may not yet be
relaxed.

\end{abstract}

\begin{keywords}
galaxies: elliptical and lenticular, cD -- galaxies: kinematics and dynamics 
\end{keywords}



\section{Introduction}
In the hierarchical scenario with $\Lambda$ Cold Dark Matter (CDM) cosmology, galaxies and clusters
form by gradually accreting other halos over time, growing into progressively larger systems. The
most massive halo eventually evolves to contain the brightest galaxy in the system, residing at the
center of the cluster potential well and/or local density peaks \citep{Beers1983}. 

This Brightest Cluster Galaxy (hereafter BCG) is often a giant
elliptical of the D or cD type (although not all BCGs are of this type). The BCG is often peculiar in terms of brightness, prevalence
of AGN activity, colours, etc:
(e.g., \citealt{Tremaine1977,Lin2010,Hearin2013,Shen2014,Skibba2006,Skibba2007,vandenBosch2007,vandenBosch2008,Skibba2009,Vulcani2014}).

One of the most unique properties of BCGs is the observed tendency for their major axes to share the same orientation as their host cluster
\citep{Sastry1968,Binggeli1982, Niederste2010,Biernacka2015}. 
This is believed to be a relic of their formation history, as it is found even at high redshifts both in observations \citep{Li2013,West2017} and cosmological  hydrodynamical simulations \citep{Okabe2020a,Ragone2020}.
The conventional explanation is that 
the BCG lies at the centre of the forming cluster halo and accretes galaxies along a preferential 
direction (collimated infall) coinciding with the major accretion filament feeding the cluster growth within the cosmic web 
(e.g., \citealt{West1994,Dubinski1998} see also \citealt{Donahue2016,Okabe2020a,Okabe2020b,Ragone2020}). This direction is then `imprinted' 
on the shape of the growing BCG halo (and the tidal debris forming the intracluster light --
\citealt{Kluge2019}) as well as that of the cluster in which it resides.
Regarding cluster major mergers, it has been shown in a recent work \citep{Ragone2020} that their frequency and geometry affect differently to the BCG alignment.
Clusters that, after a major merger, are let to evolve without further major accretions are able to restore their alignments.
Moreover, mergers along the cluster elongation axis can cause that, at the end of the accretion event, an even stronger alignment is developed. 

These observations broadly support the `central galaxy paradigm' defined by \cite{vandenBosch2005},
in which the brightest galaxy lies at the centre of the dark matter halo. However, there is now evidence
that this might not always be the case. Observations by \cite{vandenBosch2005, Skibba2011} and
\cite{Lange2018} show that the brightest halo galaxy is displaced from the halo centre (as measured by 
the X-ray peak) in up to $\sim 40\%$ of cases. In cluster environments, several authors
find significant displacements between the position of the BCG and the dark matter
halo \citep{Sanderson2009,Zitrin2012,Hikage2013,Lauer2014,Martel2014,Oliva2014,Wang2014,Hoshino2015,Rossetti2016,
Lopes2018, Zenteno2020}, while \cite{Coziol2009} detect significant offsets from the cluster mean (peculiar velocities) in
velocity space for about 1/3 of BCGs.

These off-centre BCGs are unexpected: the brightest (and most massive) cluster galaxy should occupy the
centre of its halo and be at rest with respect to the X-ray gas and the velocity distribution of the 
cluster galaxies. Because of dynamical friction, a BCG should quickly fall to the cluster centre: indeed
the observed fractions of off-centre BCGs are a factor of 2--3 higher than the predictions from the
simulations of \cite{Croton2006, Monaco2007} and \cite{DeLucia2007}. The strength of these offsets is
consistent with the bulk velocity seen in dark matter simulations \citep{Behroozi2013}: central galaxies 
may define the bottom of the potential well better than the dark matter halo \citep{Beers1983,Cui2016,Guo2016,Ye2017}.

One possibility is that the BCG is actually not
at rest with respect to the cluster centre 
(e.g., it may have been displaced by mergers or
recently infallen), another is that the dark matter halo may not be in equilibrium. There 
are known cases where the BCG does not appear
to be at rest with respect to the frame defined by the other galaxies \citep{Barbosa2018}
and examples where the X-ray gas appears to be
sloshing \citep{Markevitch2001,Churazov2003,Johnson2012,Harvey2017}. However, we expect 
that dynamical friction should quickly damp any
oscillation of the BCG in a relaxed dark matter
halo, whereas a non-relaxed halo should produce
peculiar velocities of the order of 20--30\% of
the velocity dispersion of the dark matter halo \citep{Yoshikawa2002,Ye2017}, that is broadly in agreement with the observed spatial
and dynamical offsets for BCGs.

If this is the case, however, then why are BCGs still aligned with the cluster galaxy distribution? One may naively expect that
if BCGs are moving within the cluster potential
they would not necessarily maintain their original
alignment (they could of course be displaced
mainly along the accretion axis). If BCGs
are not truly in the centre, then the origin of 
the alignment effect and its long-term preservation via collimated infall may be
difficult to explain.

In this paper we re-examine the question of BCG offsets
and we produce a comprehensive study of the displacement between BCGs and the centre of clusters' X-ray
halos. We examine whether such displaced BCGs also show peculiar velocities with respect to the cluster mean. Finally we measure the alignment effect
for BCGs at rest and offset BCGs. In the following section we describe our selection of clusters,
identification of the BCG, measurement of its position axis, measurement of the position axis of the 
X-ray halo and of the peculiar velocities. In section 3 we show our results: we assess the frequency
for BCGs to be significantly displaced from the centre of the X-ray halo and to have significant peculiar
velocity with respect to the cluster mean; we measure the alignment effect and whether it depends on the
offset from the cluster centre or mean velocity. Finally, we discuss the implications of our results in
section 4. We assume the latest cosmological parameters from the Planck Collaboration for the remainder 
of this paper \citep{Planck6}.

\section{Data}

We identified two independent cluster samples, one X-ray-selected and the other velocity-selected, as described below. 

\begin{figure}
\includegraphics[width=\columnwidth]{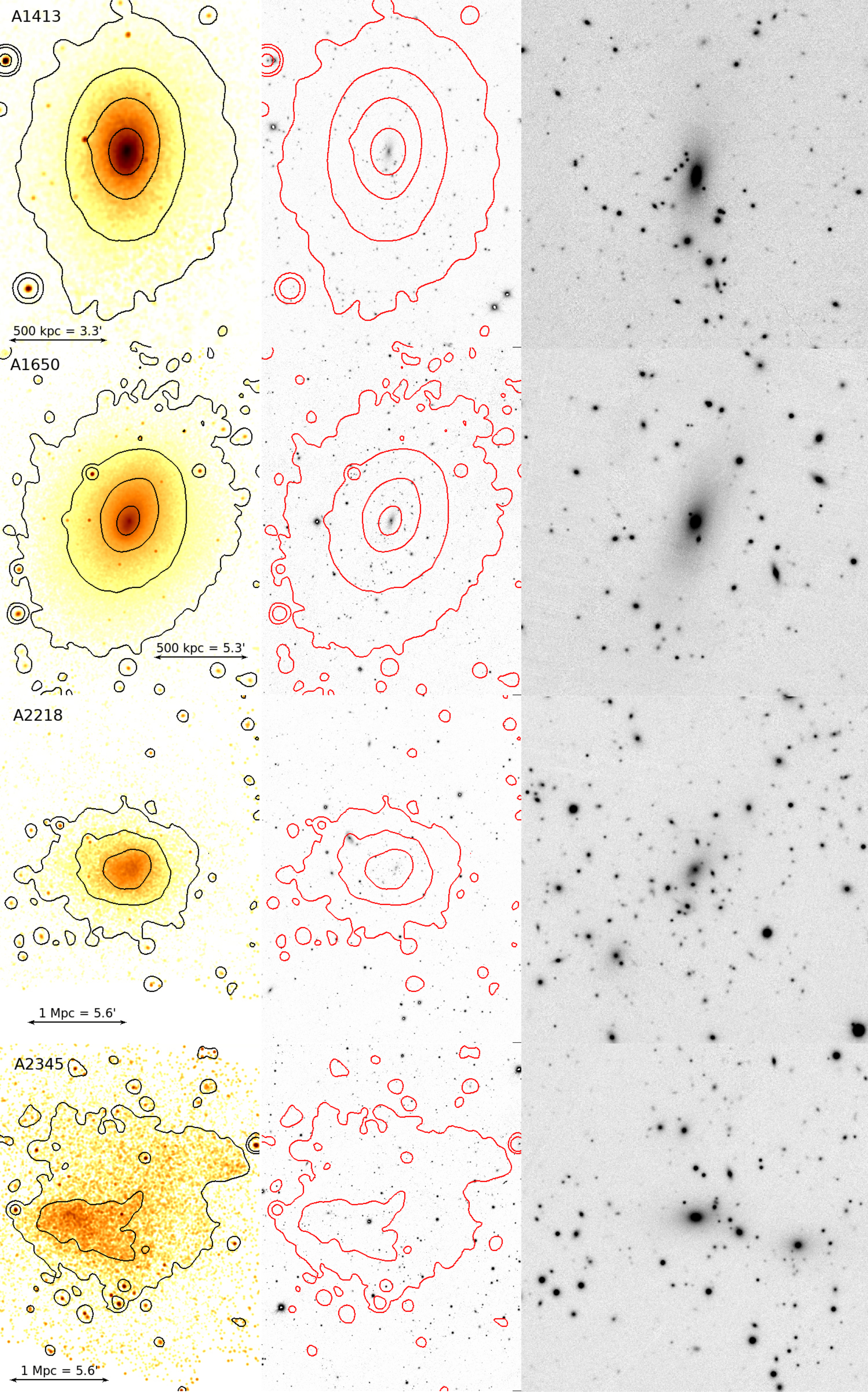}
 \caption{X-ray and optical images of several clusters from the {\it Chandra-Planck Legacy Program for Massive Clusters of Galaxies}. 
The left panels show $0.5 - 2.0$ keV, background-subtracted, exposure-map-corrected ACIS-I images from {\it Chandra}.
The middle panels show the X-ray contours overlaid on $r$-band images of the same fields from the Pan-STARRS survey.
The rightmost panels zoom into a $200^{\prime\prime} \times 200^{\prime\prime}$ 
region centered on each BCG.}
 \label{fig:clusterimages}
\end{figure}

\subsection{X-ray-selected cluster sample}

Our first cluster sample is taken from \cite{Andrade2017}, which consists of 164 clusters in the Planck
Early Sunyaev-Zel'dovich sample with $z\leqslant 0.35$ plus a flux-limited X-ray sample of 100 clusters 
with $z\leqslant 0.30$, with some overlap. All have {\it Chandra} observations obtained as part of the 
{\it Chandra-Planck Legacy Program for Massive Clusters of Galaxies}\footnote{https://hea-www.harvard.edu/CHANDRA\_PLANCK\_CLUSTERS}, with exposures that yield at least
10,000 source counts. Optical imaging is
provided by data from the Panoramic Survey Telescope and Rapid Response System (Pan-STARRS)
survey \cite{Chambers2016,Magnier2016} to identify and measure the properties of BCGs.
We choose the BCG as the brightest galaxy from
the Pan-STARRS stacked image (as in Fig. \ref{fig:clusterimages}) over a 3 Mpc region centred on the X-ray peak. In general, BCGs have distinctive appearances, dominant ellipticals often surrounded by extended halos of diffuse starlight, so identification is usually unambiguous. In cases of merging clusters or subclusters, we opted to use only the primary
component in order to focus on the most massive systems. All these clusters also have redshift information, with some overlapping with our
second sample (see below).

{\it Chandra} images were processed following standard procedures described in \cite{Vikhlinin2005} and
using calibration files CALDB 4.7.2 and $4 \times 4$ binning. Each image was further processed using the
{\it csmooth} routine in the CIAO software package \citep{Fruscione2006}, which uses the adaptive smoothing
algorithm of \cite{Ebeling2006} to create a smoothed map of the X-ray emission. Sliding cell convolution
with a Gaussian smoothing  kernel was found to be optimal for robust determination of cluster position
angles. 
A few examples are shown in Fig. \ref{fig:clusterimages}. 
The resolution of the {\it Chandra} corresponds to $\sim$ 1 to 10 kpc depending on the cluster redshifts, 
with a median resolution element of 3.5 kpc.

The projected orientation of each cluster's principal axis on the plane of the sky was determined using
Source Extractor \citep{Bertin1996}, which computes luminosity-weighted moments of the smoothed X-ray 
flux using all pixels above a 3-$\sigma$ threshold relative to the background. The cluster positions 
angles are listed in Table 1. This shows
the cluster, any other common name, the redshift, position angle of the cluster 
major axis, position angle of the BCG major axis and selection (X for X-rays and V for velocities, see below). We show the first few
lines and make the remainder of the table
available electronically. A direct comparison with position angles derived from poorer-resolution 
{\it Einstein} observations \citep{WJF1995} shows good agreement, with a median absolute difference of 
$\pm 13^{\circ}$.

\begin{table*}
\caption{Clusters and properties of BCGs (sample of full online version)}
 \begin{tabular}{||c c c c c c c||} 
 \hline
 Name & Other name & z & Cluster PA & BCG PA & $\sigma$ (km s$^{-1}$) & Selection \\ [0.5ex] 
 \hline\hline
EXO0422 &  & 0.038 & -17.8 & -9.9 & & X \\ 
Hydra A &  & 0.055 & -36.8 & -23.3 & & X \\ 
IC 1262 &  & 0.033 & 49.7 & 77.2 & & X \\ 
RXJ1958.2-3011 &  & 0.117 & -17.8 & 35.4 & & X \\ 

A7 &  & 0.106 & -44.9 & -49.2 & 1072 & V \\ 
A21 &  & 0.095 & -26.0 & -27.5 & 910 & V \\ 
A85 & G115.16-72.09 & 0.055 & -29.5/-17.6 & -29.6 & 970 & XV \\ 
XMMUJ0044.0-2033 & G106.73-83.22  & 0.292 & 30.5 & -9.0 & & X \\ 
AS0084&  & 0.108 & -77.5 & -73.2 & 717 & V \\ 
A115 & G124.21-36.48 & 0.197 & -42.6 & -32.0 & 1730 & X \\ 
A119 & G125.58-64.14 & 0.044 & 37.9/35.9 & 34.2 & 843 & XV \\ 
A133 & G149.55-84.16 & 0.057 & 24.6/32.1 & 23.2 & 700 & XV \\ 
\hline
\end{tabular}
\end{table*}

\subsection{Velocity-selected cluster sample}

Because BCGs can be offset from the cluster centre in position and/or velocity space, we compiled a 
second cluster sample by selecting all clusters with 50 or more member galaxies based on available
velocities in NED. The sample is heterogeneous but contains mainly Abell clusters plus some systems from the HeCS survey \citep{Rines2016,Rines2018} not present in the Abell catalog. This sample selection does not depend on X-ray emission, providing an independent 
check of BCG alignments when the galaxy is not at rest relative to the gravitational potential
(e.g., see \citealt{Martel2014}). As we require 50 or more spectroscopic redshifts per cluster, these are likely to be comparatively massive systems.

For all clusters we used images from PanStarrs1 to identify the brightest cluster galaxy (within
approximately the Abell radius) and take this
as the cluster center. We then retrieved all available redshifts within the $r_{200}$ radius \citep{Carlberg1997} of 
each cluster. We then used a `double gapping' method (as in \citealt{Zabludoff1990,DePropris2002}) 
to identify the velocity peak corresponding to the cluster. We sort all galaxies by velocity and 
require that the initial  sample of cluster galaxies is separated by 1000 km s$^{-1}$ gaps from 
the next galaxy in velocity space (i.e., the closest likely non cluster member).  We then compute the cluster mean velocity (location) and velocity dispersion (scale) using robust 
methods, as described by \cite{Beers1990}, using the $R$ code library \citep{R2013}. We then repeat 
our selection by requiring that the above `gaps' are 3 times this measured velocity dispersion and 
obtain the final mean velocity and velocity dispersion using the same procedure. We require that a 
minimum 50 velocities are left after the first selection. These velocity dispersions are also given in Table 1.

Cluster position angles are derived from the projected distribution of member galaxies on the plane of the sky as described in \cite{West2017}. This is done by computing the moments of inertia of the galaxy distribution. Information for this sample can be found in Table 1 as well. For clusters in common with
the X-ray selected sample the median difference in the position angle of the BCG is 15.8$^{\circ}$.

\subsection{BCG sample}

For all clusters, the BCG was identified and its properties determined using data from the Pan-STARRS PS1
survey. SDSS $r$-band images of each cluster field were downloaded as {\it fits} files from the PanSTARRS-1
database hosted at the Space Telescope Science Institute. 
Because the PS1 declination limit is $\delta \ge -30\deg$, BCGs and their host clusters at more southerly declinations were removed from the sample.
In most cases the BCG was easily identifiable 
from visual inspection of the PS1 images. In a few instances, however, multiple BCG candidates of comparable brightness 
could be seen, and so the brightest galaxy near the X-ray centroid was chosen. The sample was 
culled of any candidate BCG fainter than $M_r = 22$ to ensure that our study focuses on the most massive
galaxies. A detailed comparison of our selected BCGs with those identified in other papers   
\citep{Stott2008,Coziol2009,Lauer2014,Rossetti2016,Lopes2018} shows excellent agreement in general. 
In those few cases where there was disagreement it likely comes down to different choices among 
several plausible BCG candidates. 

Having identified the BCGs, their positions and apparent $r$-band (Kron) magnitudes were obtained from the
Pan-STARRS DR2 catalog. Each galaxy's projected distance from the X-ray centroid of its host cluster given
by \cite{Andrade2017} and its absolute magnitude, $M_r$, were calculated using the most recent cluster
redshifts in the NASA Extragalactic Database.  As expected, the BCGs have typical absolute magnitudes 
$M_r \simeq -22$ to $-23$ mag and, with few exceptions, generally reside at or near the cluster center, 
most within a few tens of kpc. Source Extractor was used to measure the projected orientation of each 
BCG's major axis and these values are listed in Table 1. 

Fig. \ref{fig:clusterimages} shows several examples of BCGs and their host clusters. Our final samples consist of 124 X-ray-selected clusters and 136 velocity-selected clusters, with 52  clusters common to both. Many of these clusters are well-known Abell clusters.

\subsection{Cosmological Hydrodynamical Simulations}
The set of cosmological hydrodynamocal simulations that we analyse in this work has already been presented in \cite{ragone2018} where it has proven to reproduce realistic BCG mass evolution histories. Furthermore, we have used it recently to assess the BCG-Cluster alignment evolution in the last 10 Gyr \citep{Ragone2020}. The simulations are similar to those presented in \cite{ragone2013}, but with an updated version of the AGN feedback scheme. 

The set consists of 29 zoomed-in Lagrangian regions evolved with a custom version of the {\footnotesize GADGET-3} code~\cite[][]{springel2005} and originally selected from a gravity-only simulation of 1 $h^{-1}$Gpc box. Part of the re-simulated regions are centered in the 24 most massive dark matter (DM) haloes of the parent cosmological volume and have masses $M_{200}\gtrsim 1.1 \times 10^{15}\, \msun$\footnote{$M_{200}$ is the mass enclosed by a sphere whose mean density is 200 times the critical density at the considered redshift. The radius of this sphere is dubbed $R_{200}$}. In addition, we randomly select 5 less massive haloes with masses $1.4 \times 10^{14} \lesssim M_{200} \lesssim 6.8 \times 10^{14}\, \msun$. Each region was re-simulated at higher resolution including hydrodynamics and all the sub-resolution baryonic physics usually taken into account in galaxy formation simulations (cooling, star formation and associated feedback, metal enrichment, AGN feedback).

The adopted cosmological parameters are: $\Omega_{\rm{m}} = 0.24$, $\Omega_{\rm{b}} = 0.04$, $n_{\rm{s}}=0.96$, $\sigma_8 =0.8$ and $H_0=72\,\kms$\,Mpc$^{-1}$. The mass resolution for the DM and gas particles is $m_{\rm{DM}} = 8.47\times10^8 \, \msunh$ and $m_{\rm{gas}} =1.53\times10^8\, \msunh$, respectively. For the gravitational force, a Plummer-equivalent softening length of $\epsilon = 5.6\, h^{-1}$\,kpc is used for DM and gas particles, whereas $\epsilon = 3\, h^{-1}$\,kpc for black hole and star particles. The DM softening length is kept fixed in comoving units for $z>2$ and in physical units at lower redshift. 
For further details on this set of simulations, we refer the reader to the above mentioned papers.

The re-simulated volumes are chosen to be large enough so that by z=0 no DM particles from the low-resolution region are found within 5 virial radii from the center of the target cluster.  For this reason, more clusters might be present in the same Lagrangian region. In particular for this work, we selected, among the two most massive clusters in each box, those  which at z=0 have at least 50 galaxies (with stellar masses $> 1\times 10^{10}\msun$). This selection criterion leads us to a sample of 38 clusters. The $M_{200}$ distribution at redshift zero has a median of $1.47\times10^{15}\msun$ and 25\% and 75\% percentiles of $6.80\times10^{14}\msun$ and $1.75\times 10^{15}\msun$, respectively. 

As for the BCGs, they are defined as the stellar particles inside 0.1 $r_{500}$ radius. This radius is similar to that at which our simulated BCGs reach a rest-frame surface brightness of $\mu_V \sim 24 {\rm~mag~arcsec^{-2}}$ \citep{ragone2018}, a classical observational value to define the galaxy limit \citep{devaucou1991}. At reshift zero, the BCGs mass distribution median is $2.16\times 10^{12}\msun$ and the 25\% and 75\% percentiles are 1.3 and  2.74 $\times 10^{12}\msun$,  respectively.

\section{Results}

\subsection{BCG offsets}

\begin{figure}
 \includegraphics[width=\columnwidth]{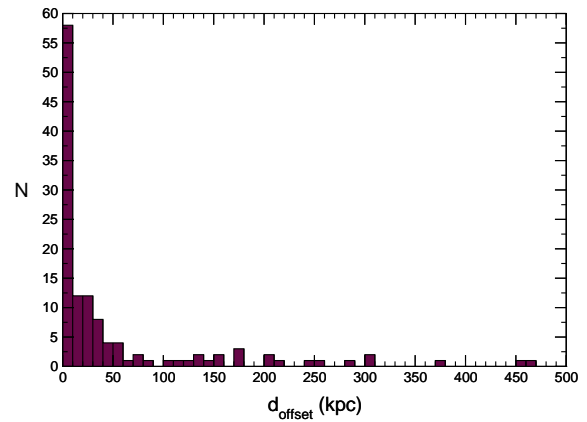}
 \caption{The distribution of projected distances between BCGs and the peak of the X-ray emission in their host clusters. Most BCGs reside at or very near the X-ray centroid. However a few are found tens or even hundreds of kpc away.}
 \label{fig:bcgoffsets}
\end{figure}

Fig. \ref{fig:bcgoffsets} shows the distribution of projected offsets between the position of the BCG and the X-ray peak. 
The great majority of BCGs in our sample lie within a few tens of kpc of the X-ray centroid -- the median separation for the X-ray-selected sample is $\sim 15$ kpc, consistent with other previous studies \citep{Lauer2014, Rossetti2016, Lopes2018} and comparable in size to the effective radii of the galaxies themselves \citep{Stott2011}. For comparison, \cite{Zitrin2012} finds distribution of displacements between the BCG and the centre of the DM halo peaked on zero, with a rms displacement of 13 kpc, a result consistent 
with our estimate above.

\begin{figure}
 \includegraphics[width=\columnwidth]{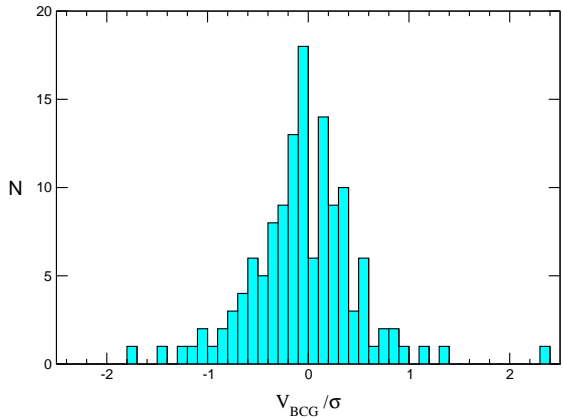}
 \caption{The distribution of BCGs peculiar velocities, defined as the difference between the BCG's radial velocity and the cluster location (the robust statistical equivalent of the cluster mean velocity) normalized by the cluster scale (the robust equivalent of the cluster velocity dispersion).}
 \label{fig:voff}
\end{figure}

Similarly, Fig. \ref{fig:voff} shows the distribution of BCG velocity offsets from the cluster mean for this sample. \cite{Martel2014}
argue that these are a more accurate measure of
true offsets than shifts from the cluster centroid in projected position on the sky.
The BCG peculiar velocities have been normalized by the cluster velocity dispersion 
(scale).
The median velocity offset for the sample is $V_{BCG}/\sigma = 0$, with 
a median {\it absolute} BCG peculiar velocity $V_{BCG}/\sigma \sim 0.26$, corresponding to typical peculiar velocities of $\sim 100$ to 200 km/s. Typical velocity errors are those of the surveys (mainly SDSS and 2dF) these velocities are largely drawn from, i.e., a few 10s of km
s$^{-1}$. These results indicate that a significant fraction of these BCGs are in motion within the cluster potential. Our estimate compares well
with \cite{Coziol2009} where BCGs had a median
$\Delta V / \sigma$ of 0.32, despite their
fewer velocities and less robust statistics
and the estimate for poorer clusters in
the COSMOS field by \cite{Gozaliasl2020}: 
the slightly  lower value we report comes from
our selection of clusters with larger velocity samples and our use of the median absolute
deviation methods. The displacements are
statistically significant: we carried out a
Monte Carlo simulation with 50 galaxies sampled
from a random Gaussian distribution and recovered the same mean velocity within 0.6\%
with a dispersion of 15\%.

\begin{figure}
 \includegraphics[width=\columnwidth]{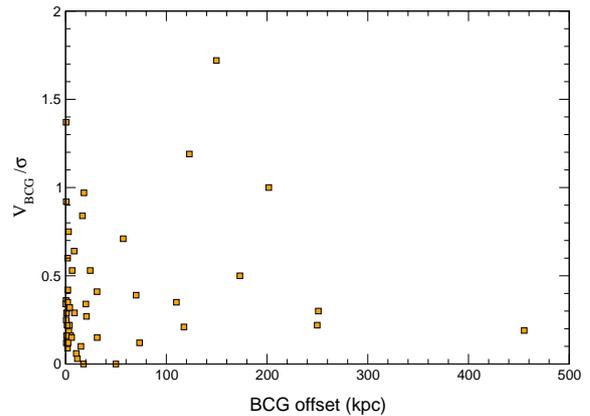}
 \caption{The distribution of BCGs peculiar velocities versus projected distance from the X-ray centroid for 52 clusters. No obvious correlation is seen.}
 \label{fig:offsetscomp}
\end{figure}

Fig. \ref{fig:offsetscomp} compares the velocity offsets with offsets from the X-ray centroid for the 52
BCGs common to both samples. No correlation is seen. This is quite surprising as, if the BCGs are displaced by mergers, one expects that shifts from the centre of the potential well also result in peculiar velocities. One possibility is that some or most of
the momentum is absorbed by the intracluster
medium. However, while most BCGs have small peculiar velocities, it is clear that a non-neglible faction are not at rest with respect to the cluster dynamical center.

\subsection{BCG alignments}
\begin{figure}
 \includegraphics[width=\columnwidth]{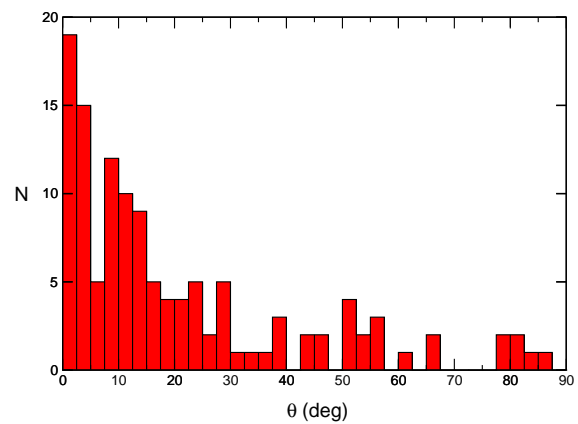}
 \caption{BCG alignments for the X-ray-selected clusters. Here $\theta$ is the acute angle between the projected major axis of each galaxy and that of the cluster
in which it resides. If the galaxy and cluster axes are perfectly aligned then $\theta = 0^{\circ}$, while random
galaxy orientations will produce a uniform distribution between $0^{\circ}$ and $90^{\circ}$. The BCGs exhibit a strong tendency to align with their host clusters.
This is confirmed by the Kuiper V, Rao and binomial statistical tests, which all indicate a probability $p \ll 1\%$ that the observed distribution of angles is consistent with random BCG orientations.}
 \label{fig:bcgxalignments}
\end{figure}

We first examine the general tendency for BCGs to share the same orientation as their host cluster
in the X-ray selected sample. 
As Fig. \ref{fig:bcgxalignments} shows, 
a strong alignment tendency is evident.  
To assess the statistical significance of these alignments, we use three different tests for isotropy:
the Kuiper V, Rao spacing, and binomial tests (see \cite{West2017} for a description of these statistical tests).
The probability that the BCGs have random orientations
with respect to their host clusters is minuscule ($\ll 1\%$)
according to these metrics.

\begin{figure}
 \includegraphics[width=\columnwidth]{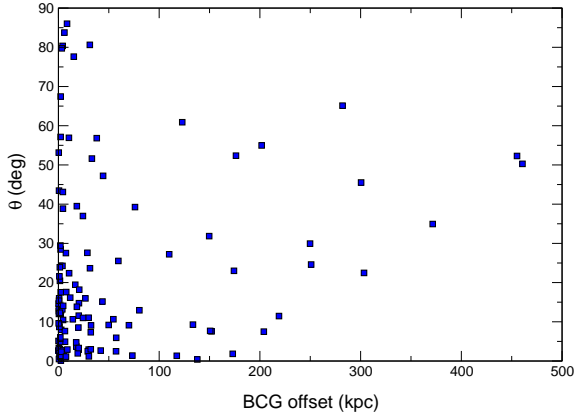}
 \caption{BCG alignments as a function of the galaxy's projected distance from the X-ray centroid of its host cluster.}
 \label{fig:bcgxalign-vs-offset}
\end{figure}

Fig. \ref{fig:bcgxalign-vs-offset} shows these same alignments as a function of BCG distance from the cluster X-ray centroid. 
Remarkably, the BCGs are aligned even when they do not reside at the cluster centre, for separations up to as much as $\sim 200$ kpc. For larger offsets, Fig. \ref{fig:bcgxalign-vs-offset} hints that the BCGs might be more randomly orientated, however no firm conclusion is possible because of the small numbers of galaxies. 

\begin{figure}
 \includegraphics[width=\columnwidth]{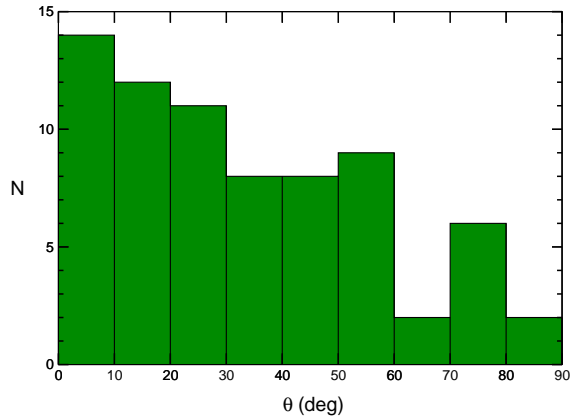}
 \caption{The direction of BCG offsets from the X-ray centroid compared to the overall cluster orientation. Here $\theta$ is the acute angle between the direction of BCG offset and the orientation of the cluster major axis. Only those clusters whose BCG is offset by more than twice the {\it Chandra} resolution are included here, a total of 72 clusters. These results reveal a tendency for the BCG to be offset preferentially along the cluster major axis.
 This is confirmed by 
 the Kuiper V, Rao, and binomial statistical tests, which all indicate a probability $p < 1\%$ that the observed distribution of offset directions is consistent with random.}
 \label{fig:slosh}
\end{figure}

In Fig. \ref{fig:slosh}, we examine the direction of BCG offsets by comparing the vector defined by the galaxy's projected position relative to the X-ray centroid with the orientation of the host cluster's major axis. There is a clear anisotropy in these offsets, with the BCGs preferentially displaced along the direction of the cluster major axis rather than in random directions.
The Kuiper and binomial tests both confirm that the distribution seen in Fig. \ref{fig:slosh} has a probability much less than $1\%$ of being consistent with 
random directional offsets, while the Rao test indicates a probability $p \sim 2\%$.

\begin{figure}
 \includegraphics[width=\columnwidth]{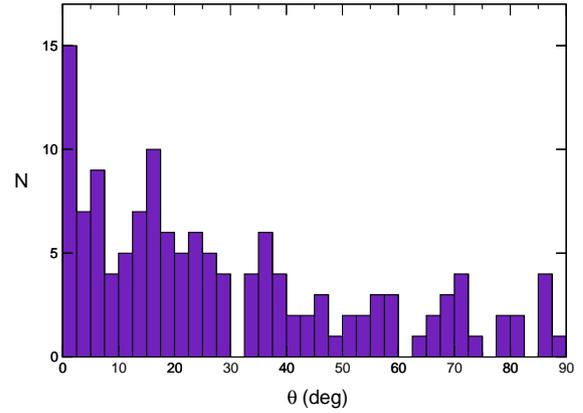}
 \caption{BCG alignments for the sample of velocity-selected clusters. As in Fig. 5, $\theta$ is the acute angle between the projected major axis of each galaxy and that of the cluster
in which it resides. Cluster position angles were determined from moments of inertia of 
the projected galaxy distribution.
The Kuiper V, Rao, and binomial statistical tests all indicate a probability $p \ll 1\%$ that the observed distribution of angles is consistent with random BCG orientations.}
 \label{fig:bcgvalignments}
\end{figure}

\begin{figure}
 \includegraphics[width=\columnwidth]{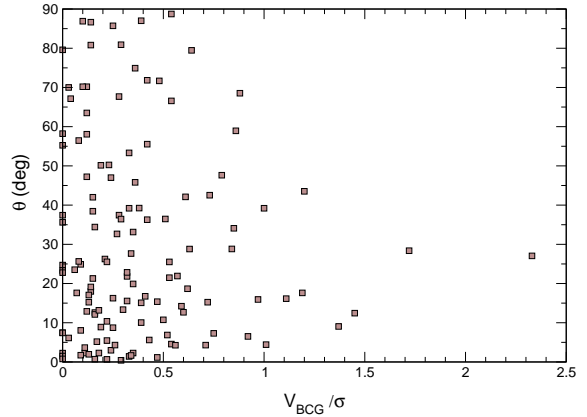}
 \caption{BCG alignments as a function of peculiar velocity. Here $\Delta V_{BCG}/\sigma$ is the absolute difference between the BCG's velocity and the mean cluster velocity, and $\sigma$ is the cluster velocity dispersion.}
 \label{fig:bcgvalign-vs-vpec}
\end{figure}

We next examine the relation between BCG peculiar velocity and alignment tendency. 
Fig. \ref{fig:bcgvalignments} shows the alignment of BCGs in the velocity-selected clusters. Again a strong  
general tendency for these galaxies to align with their host clusters is evident.
Fig. \ref{fig:bcgvalign-vs-vpec} plots the alignments as a function of BCG peculiar velocity; it appears that BCG alignments with their host clusters are largely independent of whether or not the galaxy is at rest with respect to the cluster.

\section{Discussion} 

The main findings of this paper are:

\begin{enumerate}
\item A significant fraction of BCGs have   
spatial offsets of a few tens of kpc or more from the
centroid of the X-ray light distribution that presumably 
traces the centre of the dark matter halo 
and reflects the cluster's dynamical state.
Compared with theoretical models, these
offsets are much larger than the expected
`wobble' around the centre of a standard
Cold Dark Matter halo ($< 2$ kpc vs. a few tens of kpc) as in the simulations of \cite{Kim2016} and \cite{Harvey2017}. However, the X-ray gas may not trace the dark matter halo distribution (e.g., by gas sloshing, etc.)

\item A significant number of BCGs also have significant line-of-sight peculiar velocities
relative to the cluster. 
These observations suggest that many BCGs are not currently at rest in the potential well of their host cluster. 

\item Despite the prevalence of BCG displacements, these galaxies still show a strong tendency to share the same orientation as the cluster in which they reside (cf. \citealt{Tempel2013},
a remarkably robust coherence of structures over scales from tens to thousands of kpc, as suggested by the simulations of \cite{Rhee2017}
\end{enumerate}

\begin{figure*}
\centering
\includegraphics[width=8cm]{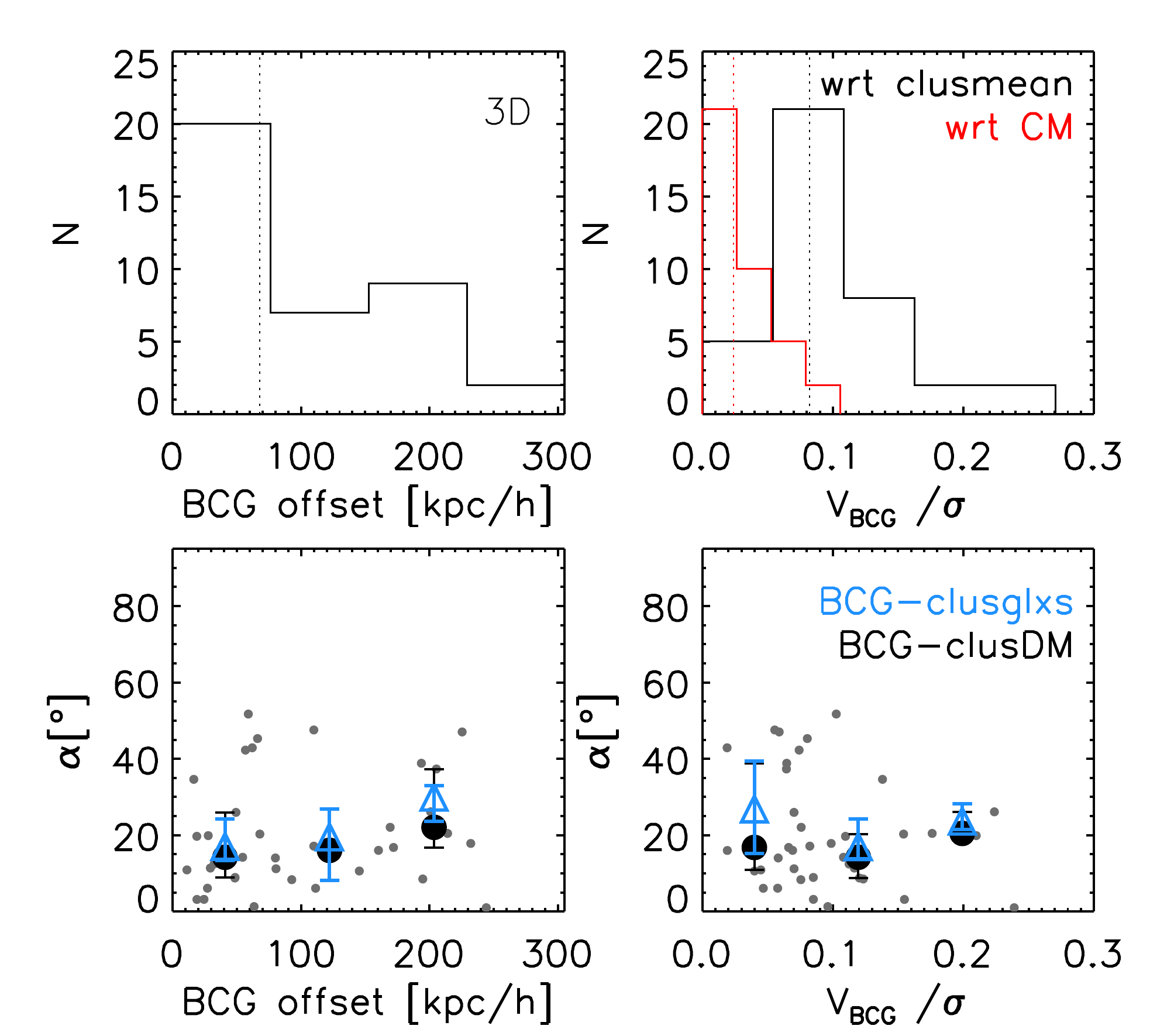}
\includegraphics[width=8cm]{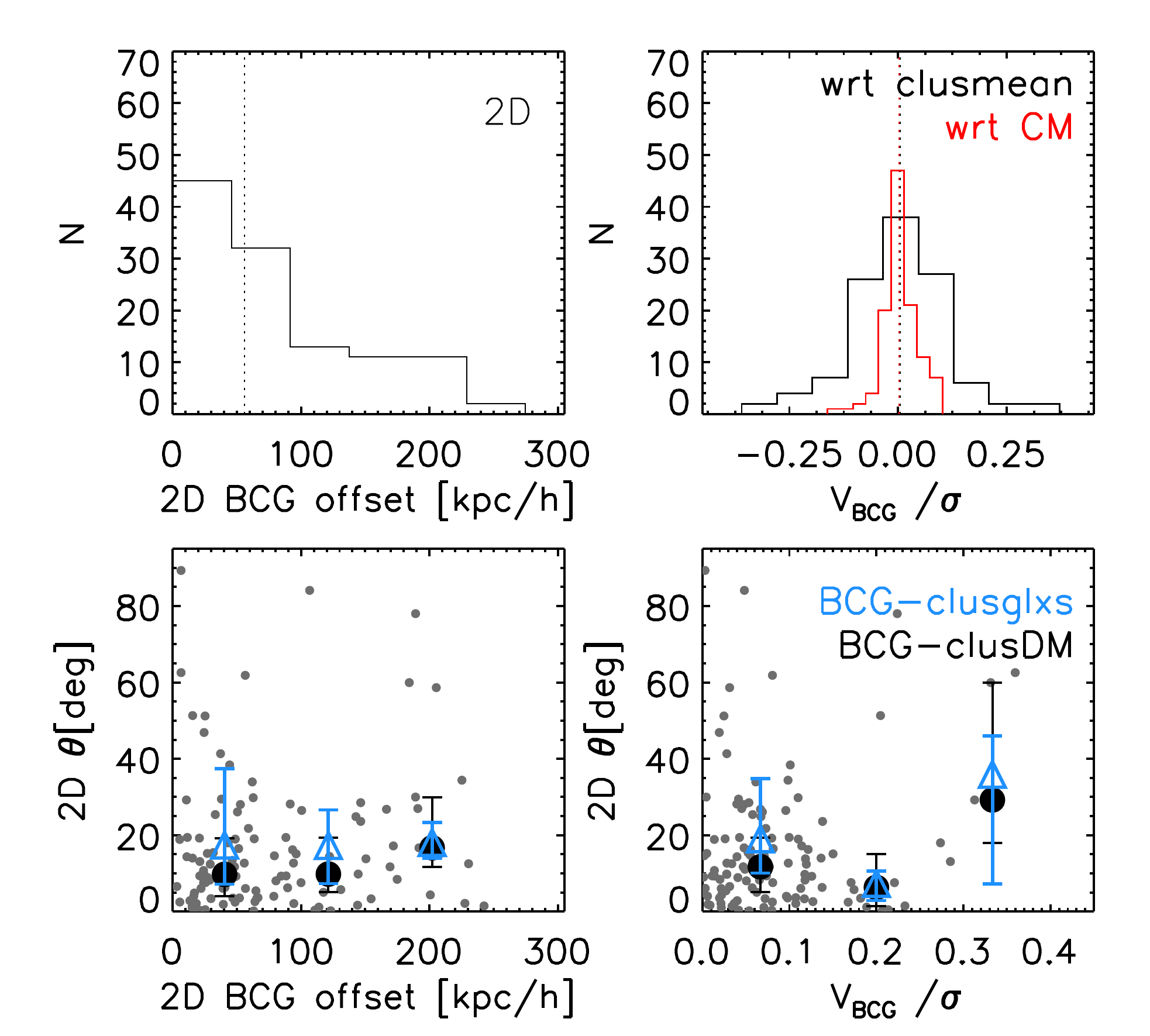}
\caption{{The four left-hand panels correspond to full 3D position and velocity space: Top panels show the distributions and medians of BCG offsets from the DM CM of clusters (left) and BCG velocities (BCG velocity with respect to the cluster-mean-velocity/DM-CM in black/red) normalized to its cluster velocity dispersion (right), respectively. 
Bottom panels show the 3D BCG alignment with the cluster galaxies (blue) and DM (black) as a function of the BCG offset (left) and the velocity with respect to the cluster mean (right). 
The four panels on the right-hand side show the same quantities but in projected space and 1D radial velocities. Here, the 114 dots are obtained from the original 38 clusters performing projections along the 3 Cartesian axes. The median of the |1D ${\rm V_{BCG}/\sigma}$| distribution is $\sim$0.07 and 0.02 depending on whether the BCG velocity is computed with respect to the cluster mean or the cluster DM CM, respectively. Large dots and bars in all panels correspond to medians and 25\%-75\% percentiles, respectively, per bin of the corresponding offset.}}
\label{fig:rf3}
\end{figure*}

\begin{figure}
 \includegraphics[width=\columnwidth]{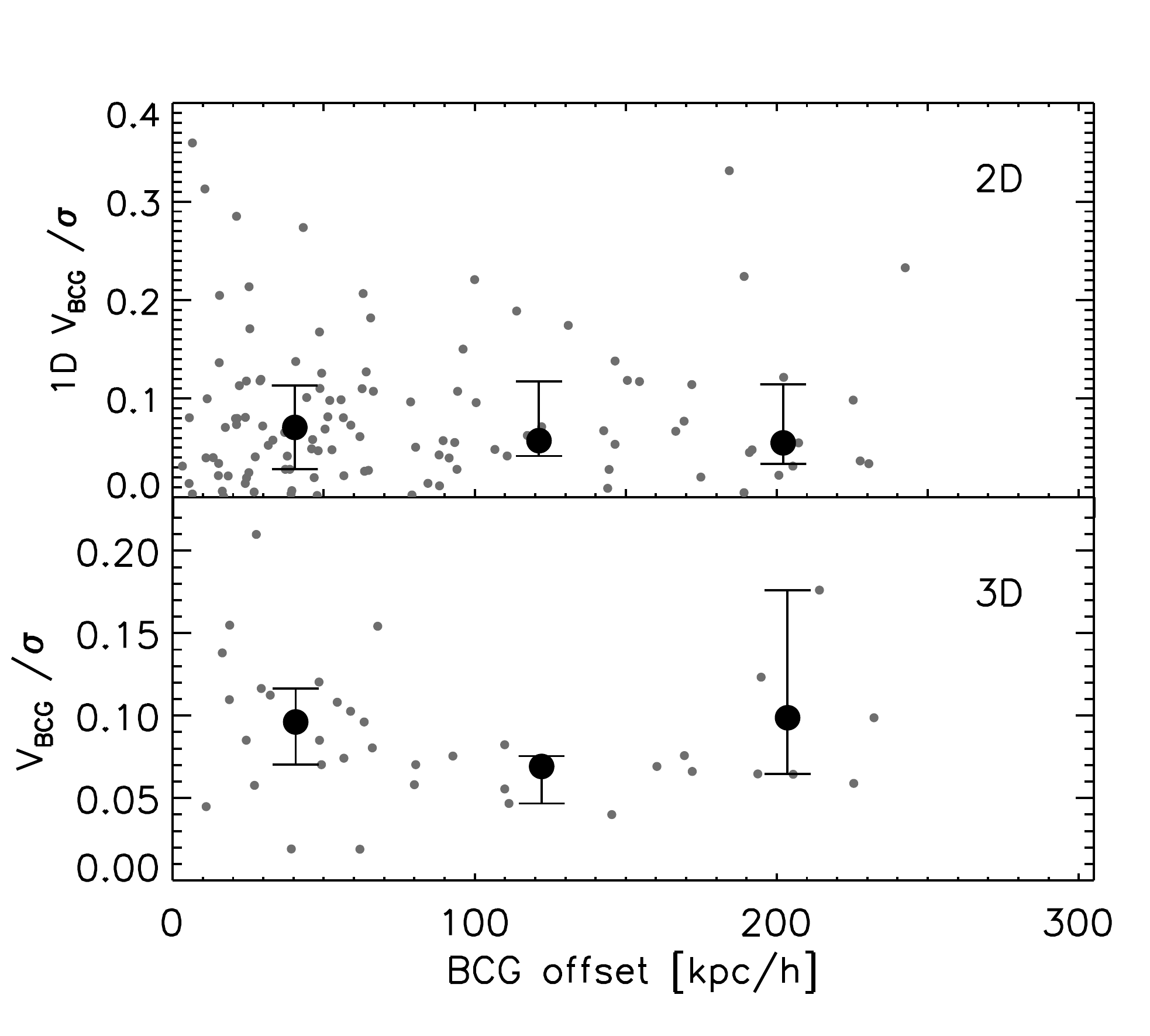}
 \caption{Top panel shows the dependence of the |1D ${\rm V_{BCG}/\sigma}$| on the projected BCG offset. The same in bottom panel but for the full 3D quantities. BCG velocities are computed with respect to the cluster mean.}
 \label{fig:offset_vel}
\end{figure}

\begin{figure}
 \includegraphics[width=\columnwidth]{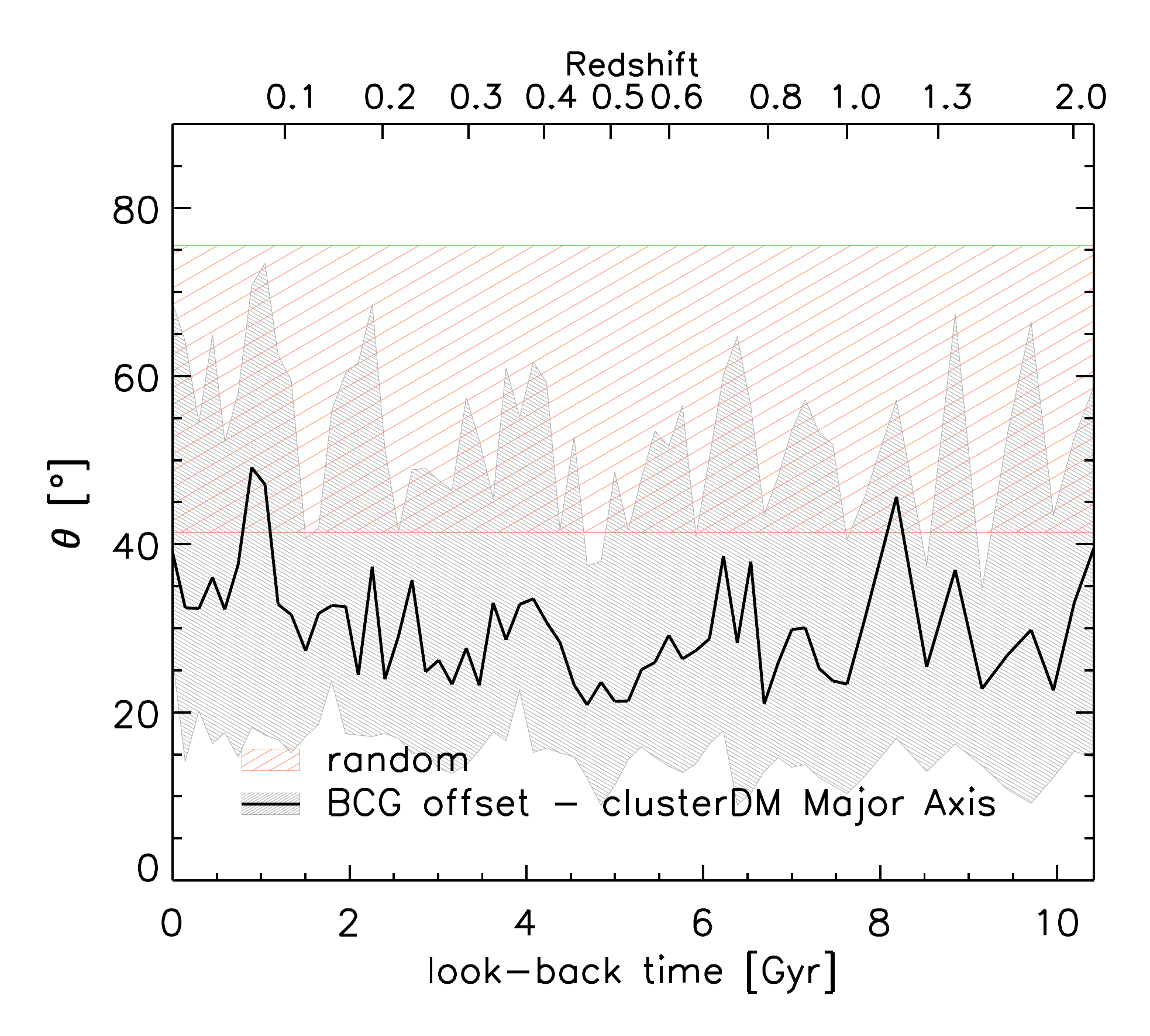}
 \caption{The direction of the BCG offsets from the clusterCM is not randomly oriented with respect to the cluster elongation in the last $\sim$ 10 Gyr. Black solid line shows, as a function of look-back time, the median 3D angle between the direction of the BCG offset from the cluster DM center of mass and the direction of the cluster DM major axis. {If these two directions were randomly oriented then it is expected a median $\theta = 60^\circ$ and 25\% 75\% percentiles of $\sim 41.4^\circ$ and $\sim 75.5^\circ$, respectively}}
 \label{fig:offset_evo}
\end{figure}

\subsection{Comparison of Observations with Simulations}

The computation of the cluster and BCG elongation axes is done as in \cite{Ragone2020}. They are obtained from the principal axes of the ellipsoids that best describe the corresponding distribution of matter. For the purposes of this work cluster principal axes and center of mass are obtained using dark matter particles within $r_{200}$. While for obtaining BCG principal axes we use star particles inside 0.1 $r_{500}$.

Fig.~\ref{fig:rf3} shows the distributions of BCG offsets in position and velocity {as well as their correlations with the BCG-Cluster alignment angle}, in full 3D space (4 left-hand panels) and in projection and radial velocity (4 right-hand panels). We find in simulations galaxies with significant offsets, in projected space the mean(median) offset is $\sim$ 53kpc(78kpc), which is larger that the mean found in the observational data ($\sim$15kpc). Conversely, the mean(median) normalized 1D BCG velocity $|V_{BCG}/\sigma|$ (with respect to the cluster mean) is $\sim$ 0.09(0.07), which is lesser than in the data (0.26).
The maximum $\Delta V/\sigma$ in these simulations is 
lower than that observed because of the smaller number of simulated cluster samples. 

BCGs in these simulations also have a tendency to be aligned with the cluster major axis. Indeed, in \cite{Ragone2020} it was shown that the signal of alignment is present since z $\lesssim$ 1.5. There is broad agreement between these simulations and the observational data on this regard. 
Namely, the alignment is still significant (within $\pm 20^{\circ}$) for off-centre BCGs and shows no dependence on the BCG  velocity with respect to the cluster mean. As shown in Fig. \ref{fig:offsetscomp} for the observations, Fig. \ref{fig:offset_vel} reveals no evident correlation between BCGs offsets and velocities.

In order to further analyse the persistence of the alignment, we reconstruct the evolutionary path of each one of the 38 clusters. We follow back in time the cluster main progenitor from $z=0$ to $z=2$ using 60 simulation outputs. We find that in the studied redshift range simulated BCG offset directions are not randomly oriented. Fig. \ref{fig:offset_evo} shows the median of the angle between the BCG offset and the cluster DM elongation axis as a function of time. This fact together with the persistence of the alignment in off-centre BCGs, suggest that BCGs lay preferentially along the cluster major axis. The cluster's major axis defines the direction of least resistance of BGC motion and the direction along which mergers preferentially occur. Mergers (if they are the cause of the BCG offsets) may therefore take place along a preferential accretion direction, as in the collimated infall models of
\cite{West1994} and \cite{Dubinski1998}.


\section{Conclusions}

The observations suggest that even in very
high mass halos the central galaxy paradigm
does not hold in a large fraction of cases,
much larger than one would expect from the
predictions of numerical codes \citep{vandenBosch2005}. In some cases, this
may be due to the mis-identification of 
a satellite as the central galaxy as the
former is not necessarily less bright or
less massive than the latter. However, galaxies
with position and/or velocity offsets cannot
all be explained in this fashion. There are
two possibilities: the BCG is still moving
within the cluster potential towards the
bottom of the potential well or the halo is
unrelaxed and oscillates. 

The persistence of the alignment effect 
even for offset galaxies would tend to 
support the latter scenario where the 
halo is not relaxed and the BCG is at rest
with respect to all other galaxies. This 
is borne out by observations for most
galaxies. However, the persistence of
alignments even for galaxies with a 
velocity offset,
at least until the peculiar velocity is
$< 40\%$ of the velocity dispersion, 
is unexpected, as in this case the 
BCG is not moving with the rest of 
the galaxies as in the non-relaxed halo
picture. One possibility is that these
are cases where the BCG is displaced by
a recent collision or cluster merger,
and that these collisions occur preferentially along
the dominant accreting filament. 

Another possibility is that the BCG is
actually moving around a constant density
core rather than a CDM cusp
\citep{Kim2016,Harvey2017}: the significant
peculiar velocities would tend to support
this. The BCG may slosh around such a core 
for long periods after a cluster merger,
although the persistence of the alignment
effect may be more difficult to maintain
in this case.

\section*{Acknowledgements}
This research has made use of the NASA/IPAC Extragalactic Database (NED),
which is operated by the Jet Propulsion Laboratory, California Institute of Technology,
under contract with the National Aeronautics and Space Administration.
The scientific results reported in this article are based in part or to a significant degree on data obtained from the Chandra Data Archive.
The Pan-STARRS1 Surveys (PS1) and the PS1 public science archive have been made possible through contributions by the Institute for Astronomy, the University of Hawaii, the Pan-STARRS Project Office, the Max-Planck Society and its participating institutes, the Max Planck Institute for Astronomy, Heidelberg and the Max Planck Institute for Extraterrestrial Physics, Garching, The Johns Hopkins University, Durham University, the University of Edinburgh, the Queen's University Belfast, the Harvard-Smithsonian Center for Astrophysics, the Las Cumbres Observatory Global Telescope Network Incorporated, the National Central University of Taiwan, the Space Telescope Science Institute, the National Aeronautics and Space Administration under Grant No. NNX08AR22G issued through the Planetary Science Division of the NASA Science Mission Directorate, the National Science Foundation Grant No. AST-1238877, the University of Maryland, Eotvos Lorand University (ELTE), the Los Alamos National Laboratory, and the Gordon and Betty Moore Foundation.

M.J.W. thanks the Finnish Centre for Astronomy with ESO (FINCA) and University of Turku for their support and hospitality during this research.

CRF acknowledges funding from the Consejo Nacional de Investigaciones Cient\'ificas y T\'ecnicas de la Rep\'ublica Argentina (CONICET)  and from the European Union's Horizon 2020 Research and Innovation Programme under the Marie Sklodowska-Curie grant agreement No 734374. 

ER acknowledges funding under the agreement ASI-INAF N.2017-14-H.0

WF and CJ acknowledge support from the Smithsonian Institution and the Chandra High Resolution Camera Project through NASA contract NAS8-03060.

We acknowledge the computing centre of INAF-Osservatorio Astronomico di Trieste, under the coordination of the Calcolo HTC in INAF - Progetto Pilota (CHIPP) \citep{bertocco2019,taffoni2020}, for the availability of computing resources and support

This work was supported by institutional research funding  \mbox{PUTJD907} of the Estonian Ministry of Education and Research.

Note added: We want to point out that comparable BCG velocity offsets were previously identified by Lauer et al (2014), who also found a correlation between BCG morphology and displacement.

\section*{Data Availability}
The data underlying this article are available in the article and in its online supplementary material.




\bibliographystyle{mnras}
\bibliography{references} 



\appendix


\bsp	
\label{lastpage}
\end{document}